\documentclass[11pt]{iopart}
\usepackage{bm}
\usepackage{hyperref}
\usepackage{amssymb}
\usepackage{amsopn}
\usepackage{feynmp}
\usepackage{deltandiagrams}

\newcommand{\ie}{\emph{i.e.}}

\newcommand{\header}[1]{{\textit{{#1}}}}

\newcommand{\R}{\mathcal{R}}
\newcommand{\fnl}{f_{\mathrm{NL}}}
\newcommand{\tnl}{\tau_{\mathrm{NL}}}
\newcommand{\cn}{\mathrm{\mathsf{c}}}
\newcommand{\powerspectrum}{\mathcal{P}}
\newcommand{\Mfact}{\mathcal{M}}
\newcommand{\Z}[2]{\vect{Z}_{{#1}{#2}}}
\newcommand{\ktot}{k_t}

\renewcommand{\d}{\mathrm{d}}
\newcommand{\vect}[1]{\bm{\mathrm{{#1}}}}
\renewcommand{\e}[1]{\mathrm{e}^{{#1}}}

\newcommand{\imag}{\mathrm{i}}
\newcommand{\reals}{\bm{\mathrm{R}}}

\newcommand{\grad}{\nabla}
\renewcommand{\leq}{\leqslant}
\renewcommand{\geq}{\geqslant}
\DeclareMathOperator{\gradient}{grad}
\DeclareMathOperator{\divergence}{div}
\DeclareMathOperator{\curl}{curl}

\begin{document}
\title{The inflationary trispectrum}
\date{\today}
\author{David Seery,$^1$ James E. Lidsey$^1$ and Martin S. Sloth$^2$}
\address{$^1$ Astronomy Unit, School of Mathematical Sciences\\
  Queen Mary, University of London\\
  Mile End Road, London E1 4NS\\
  United Kingdom}
\vspace{2mm}
\address{$^2$ Department of Physics and Astronomy\\
  University of Aarhus\\
  Ny Munkegade, Building 1520\\
  DK-8000 Aarhus C\\
  Denmark}
\eads{\mailto{D.Seery@qmul.ac.uk}, \mailto{J.E.Lidsey@qmul.ac.uk},
      \mailto{sloth@phys.au.dk}}
\submitto{JCAP}
\begin{abstract}
We calculate the trispectrum of the primordial curvature perturbation
generated by an epoch of slow-roll inflation in the early universe, and
demonstrate that the non-gaussian signature imprinted at horizon crossing
is unobservably small, of order $\tnl \lesssim r/50$,
where $r < 1$ is the tensor-to-scalar ratio.
Therefore any primordial non-gaussianity observed in future
microwave background experiments is likely to have been
synthesized by gravitational effects on superhorizon scales.
We discuss the application of Maldacena's consistency condition to the
trispectrum.

\vspace{3mm}
\begin{flushleft}
  \textbf{Keywords}:
  Inflation,
  Cosmological perturbation theory,
  Physics of the early universe.
\end{flushleft}
\end{abstract}
\maketitle

\section{Introduction}

In the inflationary scenario
\cite{Starobinsky:1980te,Sato:1980yn,Guth:1980zm,
Hawking:1981fz,Albrecht:1982wi,
Linde:1981mu,Linde:1983gd},
the primordial curvature perturbation,
$\zeta$, is generated via the vacuum fluctuations of
one or more light scalar fields 
\cite{Bardeen:1983qw,Guth:1982ec,Hawking:1982cz,Hawking:1982my}. 
It is the only relevant perturbation in the simplest class of  
single-field models. More generally, however, $\zeta$ is sensitive to the
model under consideration.
It may be accompanied by correlated or uncorrelated isocurvature perturbations
\cite{Langlois:2000ar,Lyth:2003ip} which source its evolution,
or it may be generated
after the end of inflation by the decay of another field
\cite{Enqvist:2001zp,Lyth:2001nq,Moroi:2001ct,Mollerach:1989hu}.

It has recently become clear that interesting information 
about the particle physics that drove inflation 
is encoded in the non-gaussian features of the curvature perturbation 
\cite{Bartolo:2004if,Maldacena:2002vr,Rigopoulos:2004gr,Rigopoulos:2004ba,
Seery:2005gb,Rigopoulos:2005xx,Lyth:2005fi,Lyth:2005qj,Zaballa:2006pv,
Allen:2005ye,Alabidi:2005qi,Gupta:2002kn,Gupta:2005nh,
Lyth:2002my,
Boubekeur:2005fj,Lyth:2005du,Lyth:2006gd,Malik:2006pm,Enqvist:2005pg,
Enqvist:2004ey,Huang:2005nd,Barnaby:2006cq,Valiviita:2006mz}.
Although observations of the Cosmic Microwave Background (CMB) 
anisotropy constrain the spectrum of perturbations to 
be close to scale-invariance with very nearly gaussian statistics
\cite{Vielva:2003et,McEwen:2004sv,Tojeiro:2005mt,McEwen:2006yc}, 
some non-gaussianity is inevitable due to the universal coupling of 
all matter fields to gravity
\cite{Bartolo:2001cw,Bartolo:2003gh,Acquaviva:2002ud}. 
The question remains, however, as to the 
precise level of non-gaussianity that is generated. 
This is sensitive to details of the inflationary model, such as the number
of fields which contribute to the energy density of the universe during
inflation \cite{Vernizzi:2006ve,Kim:2006te,Alabidi:2006hg,Enqvist:2004bk,
Piao:2006nm}, and also to possible non-canonical structures in the 
scalar field Lagrangian 
\cite{Seery:2005wm,Chen:2006nt,Alishahiha:2004eh,Arkani-Hamed:2003uy,
Creminelli:2003iq,Calcagni:2004bb,Seery:2006tq}.

If pure gaussian statistics are respected, 
any $n$-point correlator of the curvature perturbation
$\langle \zeta \cdots \zeta \rangle$ can always be expressed in terms 
of the two-point correlator $\langle \zeta \zeta \rangle$. 
A breakdown of this rule for any $n \geq 3$ indicates  
a departure from gaussianity. Since some higher-order correlations 
may be more easily detectable
\cite{Aghanim:2003fs,DeTroia:2003tq,Castro:2002df}
than others, it is important
to quantify the theoretical predictions for as many values of $n$
as possible. To date, the majority of quantitative theoretical 
studies have focused on the 
three-point function, or equivalently the `bispectrum' $B_\zeta$,
\begin{equation}
  \langle \zeta(\vect{k}_1)
          \zeta(\vect{k}_2)
          \zeta(\vect{k}_3) \rangle
  = (2\pi)^3 \delta(\sum_i \vect{k}_i) B_\zeta .
\end{equation}
Precise gaussianity corresponds to $B_\zeta = 0$. 
In practise, the non-gaussianity in the three-point function 
generated in a particular model of inflation can be 
quantified in terms of a dimensionless quantity, $\fnl$, which is defined by 
\cite{Maldacena:2002vr,Komatsu:2001rj,Verde:1999ij}%
\footnote{An alternative sign convention is used in
Ref. \cite{Lyth:2006gd}.}
\begin{equation}
  \label{fnl}
  B_\zeta = - \frac{6}{5} \fnl
  [ P_{\zeta}(k_1) P_{\zeta}(k_2) + \mbox{cyclic permutations} ] ,
\end{equation}
where $P_{\zeta}(k_1)$ is the power spectrum of $\zeta$, such that  
$\langle \zeta(\vect{k}_1) \zeta(\vect{k}_2) \rangle =
(2\pi)^3 \delta( \sum_i \vect{k}_i ) P_\zeta(k_1)$. Current observational
bounds require $|\fnl| \lesssim 100$
\cite{Spergel:2006hy} and are consistent with $\fnl = 0$, 
whereas a value of $|\fnl| \sim 3$ should be detectable in the near future
\cite{Komatsu:2001rj}. On the other hand,
it is a robust prediction that the non-gaussian signal imprinted at horizon 
crossing during single-field inflation is $|\fnl| \sim 0.01$, which 
is too small to be observable,
\cite{Maldacena:2002vr}
and the situation is similar in models with more than one field
\cite{Seery:2005gb,Lyth:2005qj,Vernizzi:2006ve,
Zaballa:2006pv,Alabidi:2005qi,Kim:2006te,Alabidi:2006hg}.
(For an alternative analysis,
see Refs. \cite{Rigopoulos:2005ae,Rigopoulos:2005us}.)

In this paper, we make a comparable quantitative prediction for the
primordial four-point function, or equivalently the `trispectrum' $T_\zeta$
\cite{Okamoto:2002ik,Kogo:2006kh}:
\begin{equation}
  \langle \zeta(\vect{k}_1)
          \zeta(\vect{k}_2)
          \zeta(\vect{k}_3)
          \zeta(\vect{k}_4) \rangle
  = (2\pi)^3 \delta(\sum_i \vect{k}_i) T_\zeta .
\end{equation}
This can be quantified by defining a dimensionless `non-linearity'
parameter $\tnl$ such that \cite{Boubekeur:2005fj,Lyth:2005fi,Alabidi:2005qi}
\begin{equation}
  \label{tnl}
  T_\zeta = \frac{1}{2} \tnl [ P_\zeta(k_1) P_\zeta(k_2)
  P_\zeta(k_{14}) + \mbox{23 permutations} ] ,
\end{equation}
where $\vect{k}_{ij} = \vect{k}_i + \vect{k}_j$, and
the condition $\sum_i \vect{k}_i = 0$ ensures that one-half of the
permutations are equal to the other half, giving twelve distinct terms.

At present there is only a weak experimental bound
on the non-linearity parameter, which is
roughly $|\tnl| \lesssim 10^8$ \cite{Alabidi:2005qi}.
The WMAP satellite should strengthen this to 
$|\tnl| \lesssim 2 \times 10^4$ and the Planck satellite will be sensitive 
to a value of $|\tnl| \sim 560$  
\cite{Kogo:2006kh}.
Therefore, there is a pressing need to develop estimates of $\tnl$ in those
models of inflation which we wish to confront with experimental data in the
near future.
An expression for $\tnl$ was presented recently by 
Alabidi \& Lyth \cite{Alabidi:2005qi}
(see also \cite{Lyth:2006gd}).
However, these authors neglected any contributions arising
from quantum interactions at horizon crossing.
The purpose of the present paper is to provide a more complete expression
for the non-linearity parameter
where such quantum effects are taken into account. We consider a
general multi-field scenario in which the fields are minimally
coupled to gravity with a target space metric $\delta_{\alpha \beta}$,
and where the slow-roll approximation 
applies around the time of horizon crossing. 

The plan of this paper is as follows.
In \S\ref{sec:deltan}, we outline the $\delta N$ formalism for
computing the trispectrum of the curvature perturbation on 
super-horizon scales. In \S\ref{sec:ff}, we calculate the fourth-order
interaction lagrangian for the field perturbations, including 
both scalar \emph{and} vector contributions to 
the action for the first time. Given the form 
of this lagrangian, the four-point expectation value of the field 
fluctuations, and hence the momentum-dependence of the trispectrum,  
can be calculated by employing the Schwinger-Keldysh formalism
\cite{Schwinger:1960qe,DeWitt:2003pm,Calzetta:1986ey,Jordan:1986ug}. 
This is achieved in \S\ref{sec:expect} and represents one of the
main results of the paper. We then derive an upper bound on the 
magnitude of the trispectrum in \S\ref{sec:bound} and find that
$\tnl \lesssim r/50$, where $r < 1$ is the tensor-to-scalar
ratio. Such a result implies that any non-gaussian signature 
generated at horizon crossing from
the four-point correlator of the curvature perturbation 
will be unobservably small. We then consider the case of single-field inflation
in \S\ref{sec:consistency}, where the curvature perturbation 
is automatically conserved on superhorizon scales.
This conservation leads to a `consistency' condition between 
the correlators of $\zeta$ of the type
discussed by Maldacena \cite{Maldacena:2002vr}. 
We conclude with a discussion in \S\ref{sec:conclude}.

\section{The $\delta N$ formalism for the trispectrum}
\label{sec:deltan}

The most efficient method for calculating the four-point correlator
of the curvature perturbation 
is to employ the $\delta N$ formalism
introduced by Starobinsky \cite{Starobinsky:1986fx,Sasaki:1995aw}, and
generalized to non-linear perturbation theory by
Lyth, Malik \& Sasaki \cite{Lyth:2004gb}.%
\footnote{See also Langlois \& Vernizzi \cite{Langlois:2006vv}
for an alternative non-linear generalization.}
In this approach, it is 
assumed that the evolution of the universe in causally disconnected regions 
is like the evolution of separate locally unperturbed universes, where 
pressure and density can take different values
\cite{Wands:2000dp,Rigopoulos:2003ak,Salopek:1990jq}.
This implies that $\zeta = \delta N$
\cite{Starobinsky:1986fx,Sasaki:1995aw,Lyth:2004gb,Lyth:2005du,Lyth:2005fi},
where $N(\phi,\rho)$ is the number of e-folds of expansion between an initial
spatially flat hypersurface on which the fields have values $\phi^\alpha$
and a final uniform-density slice on which the energy density 
is given by $\rho$. On large scales, $\zeta$ is equivalent
to the comoving curvature perturbation, $\R$, up to a sign convention.

As demonstrated by Lyth \& Rodr\'{\i}guez
\cite{Lyth:2005du},
if slow-roll inflation is valid on the initial flat slice,
$\zeta$ can be written as a power series in terms of the field
perturbations at that time
\begin{equation}
  \label{zeta}
  \zeta = N_{,\alpha} \delta \phi^\alpha + \frac{1}{2}
  N_{,\alpha\beta} \delta\phi^\alpha \delta\phi^\beta + \cdots ,
\end{equation}
where $N_{,\alpha} = \partial N/\partial \phi^\alpha$ and 
$\alpha$ indexes the space of light fields
$ \phi^\alpha$. The perturbations in these 
fields are denoted by $ \delta\phi^\alpha$ and are defined on the flat 
hypersurfaces.  
The leading-order relation for the power spectrum of the curvature 
perturbation is then given by 
\begin{equation}
  \label{spectrum}
  P_{\zeta} = \delta^{\alpha\beta} N_{,\alpha} N_{,\beta} P_{\ast} ,
\end{equation}
where a subscript `$\ast$' implies that background quantities 
should be evaluated at horizon crossing and $P_{\ast}$ 
is related to the dimensionless power spectrum, $\powerspectrum_{\ast}$, 
for a massless scalar field by 
\begin{equation}
  \label{masslesspower}
  \powerspectrum_{\ast}(k) = k^3 P_{\ast}(k) / 2\pi^2
  = (H_{\ast}/2\pi)^2 .
\end{equation}
In principle, $P_\zeta$ receives corrections from higher derivatives of
$N$. We will assume that these corrections
can be neglected when computing the leading non-gaussianity. 
This is reasonable since it is known experimentally that 
$\zeta$ is dominated by a very nearly gaussian contribution. 

\begin{figure}
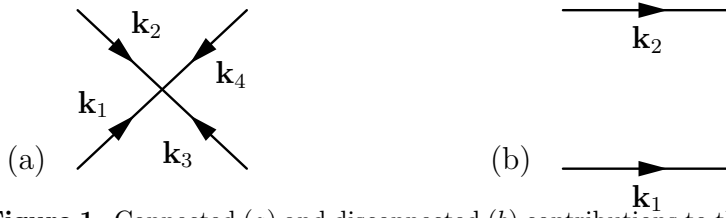

  \hfill
  (a) \quotediagram(80,60){connected.1}{connected.t1}
  \hfill
  (b) \quotediagram(80,60){connected.2}{connected.t2}
  \hfill \mbox{}
  \caption{\label{fig:connected}
  Connected $(a)$ and disconnected $(b)$
  contributions to
  the four-point function of the $\{ \delta\phi \}$.
  The disconnected contribution factorizes into a product of two copies
  of the power spectrum, and contains no information concerning
  primordial non-gaussianities beyond those encoded in loop
  corrections, which are expected to be unobservably
  small \cite{Sloth:2006az}.}
\end{figure}

To calculate $\langle \zeta \zeta \zeta \zeta \rangle$, one forms the product
of four copies of Eq.~\eref{zeta} and takes an expectation value in the initial
vacuum state \cite{Lyth:2006qz}.
The lowest-order contribution in derivatives of $N$ is given by 
\begin{equation}
  \label{fourpt-a}
  \fl
  \Delta \langle \zeta(\vect{k}_1) \zeta(\vect{k}_2)
                 \zeta(\vect{k}_3) \zeta(\vect{k}_4) \rangle =
   N_{,\alpha} N_{,\beta} N_{,\gamma} N_{,\delta}
   \langle \delta \phi^\alpha(\vect{k}_1) \delta \phi^\beta(\vect{k}_2)
           \delta \phi^\gamma(\vect{k}_3) \delta \phi^\delta(\vect{k}_4)
   \rangle ,
\end{equation}
where the correlator of field perturbations on the right-hand side is evaluated
a few e-foldings after horizon crossing, 
such that the fields have had sufficient time
to become classical but have not evolved significantly.
The $\delta\phi$-correlator in
\eref{fourpt-a} can be separated into two pieces.
The first is an irreducible connected part
in the sense of Feynman diagrams (see Figure~\ref{fig:connected}(a)), which
we denote by $\langle \delta \phi^\alpha \delta \phi^\beta
\delta\phi^\gamma \delta\phi^\delta \rangle_{\cn}$, and which is absent when
the fields perturbations are precisely gaussian. The second 
contribution is a reducible part
and is given by the sum over all ways of combining the four fields
into pairs, with each pair yielding a copy of the spectrum.
This reducible part is a \emph{disconnected} contribution, and is always
present, even when the $\delta \phi^{\alpha}$ are gaussian.
It therefore contains no more information concerning the primordial
non-gaussianity than is already present in the power spectrum, and in
the remainder of this paper we consider only connected contributions
to the trispectrum.
In principle, the contribution to
$\langle \zeta\zeta\zeta\zeta \rangle$
in Eq.~\eref{fourpt-a}
is accompanied by terms which are higher-order in the derivatives
of $N$, of which an example was computed by
Sasaki, V\"{a}liviita \& Wands \cite{Sasaki:2006kq}.
We will briefly address these contributions in
\S\ref{sec:consistency},
and will return to them in a forthcoming publication.

In obtaining a prediction for the non-linearity parameter $\tnl$ from 
Eq.~\eref{fourpt-a}, it proves very useful to decompose the trispectrum 
into a number of copies of the dimensionless power spectrum,
$\powerspectrum_{\ast}$.
For a target space metric $\delta_{\alpha\beta}$, 
one may define the connected part of the trispectrum in terms of a 
form-factor $\Mfact_4(\vect{k}_1,\vect{k}_2,\vect{k}_3,\vect{k}_4)$, 
which parametrizes the momentum-dependence of $\langle \delta\phi 
\delta \phi \delta \phi \delta \phi \rangle$. More specifically,  
combining Eqs.~\eref{spectrum},
\eref{masslesspower} and \eref{fourpt-a} leads to a   
definition for $\Mfact_4$ given by
\begin{equation}
  \label{trispect}
  T_{\delta\phi|\cn}^{\alpha\beta\gamma\delta}
  = \frac{4\pi^6}{\prod_i k_i^3 }
  \powerspectrum_{\ast}^3
  \sum_{\mbox{\scriptsize perms}} \delta^{\alpha\beta} \delta^{\gamma\delta}
  \Mfact_4(\vect{k}_1,\vect{k}_2,\vect{k}_3,\vect{k}_4) ,
\end{equation}
where $i \in \{ 1, 2, 3, 4 \}$ and the 
sum over permutations includes all
rearrangements of the indices $\{ \alpha, \beta, \gamma, \delta \}$ which
simultaneously permute $\{ \vect{k}_1, \vect{k}_2, \vect{k}_3,
\vect{k}_4 \}$. Note that the form-factor 
may depend on the relative orientation of the $\vect{k}_i$ and
not merely on their magnitude. 
Note also that the $k_i$ should have approximately 
equal magnitude in order for the epoch of 
horizon crossing to be more or less unique.

The momentum-dependence of $\tnl$ may 
then be deduced by comparing Eqs.~\eref{fourpt-a} and \eref{trispect} 
with Eq.~\eref{tnl}. This is facilitated by rewriting Eq.~\eref{tnl} in the
form
\begin{equation}
  \label{tnl-def}
  T_{\zeta|\cn} = \frac{4\pi^6}{\prod_{i \in \{ 1, 2, 3, 4 \}} k_i^3} \tnl
  (\delta^{\alpha\beta} N_{,\alpha} N_{,\beta})^3 \powerspectrum_\ast^3
  \sum_{i < j} \sum_{m \neq i, j} k_i^3 k_j^3 (k_{im}^{-3}
  + k_{jm}^{-3}) ,
\end{equation}
where we have substituted for the power spectrum \eref{spectrum}. 
It then follows from Eqs.~\eref{fourpt-a},
\eref{trispect} and~\eref{tnl-def} that the contribution
to $\tnl$ that is generated at horizon crossing is given by 
\begin{equation}
\label{contribution}
  \Delta \tnl = \frac{\sum_{\mbox{\scriptsize perms}}
  \Mfact_4(\vect{k}_1,\vect{k}_2,\vect{k}_3,\vect{k}_4)}
  {(\delta^{\alpha\beta}N_{,\alpha} N_{,\beta})
   \sum_{i < j} \sum_{m \neq i, j} k_i^3 k_j^3 (k_{im}^{-3} +
   k_{jm}^{-3})} ,
\end{equation}
where the sum is over all permutations of the $\{ \vect{k}_1, \cdots,
\vect{k}_4 \}$.

Finally, the non-linearity parameter can be 
related to the `tensor-to-scalar' ratio, $r$, which is defined in terms of 
the corresponding power spectra such that  
\begin{equation}
  r = \frac{8 \powerspectrum_\ast}{\powerspectrum_\zeta} =
  \frac{8}{\delta^{\alpha\beta} N_{,\alpha} N_{,\beta}} ,
\end{equation}
where Eq.~\eref{spectrum} has been employed.
Hence, it follows that
\begin{equation}
  \label{tnlr}
  \Delta \tnl = 
  \frac{r}{8} \frac{\sum_{\mbox{\scriptsize perms}} \Mfact_4}{\sum_{i < j}
  \sum_{m \neq i, j} k_i^3 k_j^3 (k_{im}^{-3} + k_{jm}^{-3})} .
\end{equation}
The magnitude of this contribution can therefore 
be derived immediately  
once the form-factor $\Mfact_4(\vect{k}_1,\vect{k}_2,\vect{k}_3,
\vect{k}_4)$ has been calculated \cite{Lyth:2005qj}. This 
calculation will be the focus of the following two sections. 

\section{The fourth-order action for the perturbations}
\label{sec:ff}

The appropriate formalism for computing expectation values such as
$\langle \delta\phi^\alpha \delta\phi^\beta \delta\phi^\gamma
\delta\phi^\delta \rangle$ is the so-called `closed time path' integral,
which was developed by Schwinger \cite{Schwinger:1960qe} and others
\cite{DeWitt:2003pm,hajicek},
and subsequently extended to cosmology \cite{Calzetta:1986ey,Jordan:1986ug,
Maldacena:2002vr,Weinberg:2005vy,Weinberg:2006ac,
Seery:2006tq,Seery:2006wk}. 
This integral expresses the four-point correlator as
\begin{eqnarray}
  \fl\nonumber
  \langle \delta\phi^\alpha(t',\vect{k}_1) \delta\phi^\beta(t',\vect{k}_2)
          \delta\phi^\gamma(t',\vect{k}_3) \delta\phi^\delta(t',\vect{k}_4)
  \rangle
  = \\ \label{ctp}
  \int [\d \delta\phi_+ \, \d \delta\phi_-] \;
  \delta\phi^\alpha_+(t',\vect{k}_1) \delta\phi^\beta_+(t',\vect{k}_2)
  \delta\phi^\gamma_+(t',\vect{k}_3) \delta\phi^\delta_+(t',\vect{k}_4)
  \;
  \exp \imag W ,
\end{eqnarray}
where $W$ is a weight involving the action for the
field perturbations $\delta\phi^\alpha$ and takes the form
\begin{equation}
  \label{weight}
  W[\delta\phi_+,\delta\phi_-] =
  \int_{-\infty}^{t_{\ast}} \d t \, \d^3 x \;
  L(\delta\phi_+) -
  \int_{-\infty}^{t_{\ast}} \d t \, \d^3 x \;
  L(\delta\phi_-) ,
\end{equation}
where $L(\delta\phi)$ is the lagrangian for the $\{ \delta \phi \}$.
In Eqs.~\eref{ctp}--\eref{weight},
we have adopted a common time of observation $t'$, and $t_\ast$ is any
fiducial time satisfying $t' < t_\ast$. 
The functional integral is over all `forward-going' fields
$\delta\phi_+$
\cite{hajicek} which begin in the vacuum state at $t \rightarrow -\infty$
together with all `backward-going' fields $\delta\phi_-$ which obey the same
boundary
condition and coincide with the $\delta\phi_+$ at $t_\ast$.
This differs from the conventional Feynman expression 
in which only the forward-going field is present
\cite{Weinberg:1995mt,Peskin:1995ev}. This difference arises 
because the Feynman formula computes a scattering
amplitude from the past to the future, whereas the Schwinger 
formula computes an expectation value.

The lagrangian can be naturally expressed as a sum of terms involving a 
definite power of the perturbation $\delta\phi$,
\begin{equation}
  L(\delta\phi) = \sum_{k=2}^{\infty} L_k(\delta\phi)
\end{equation}
where $L_k$ contains $k$ such powers. Hence, the gaussian free theory is 
completely specified by $L_2$, but the $L_k$ terms with $k \geq 3$ 
correspond to interactions.
In the perturbative r\'{e}gime these can be assumed to be small corrections
to the free theory, and generate the
non-gaussianities we wish to calculate. In order to compute the $n$-point
expectation value, it is necessary to know the functional 
forms for all $L_k$ with $k \leq n$.
The $L_2$ and $L_3$ terms were obtained
in Ref.~\cite{Seery:2005gb} and are sufficient to calculate the three-point
expectation value and therefore the bispectrum $B_{\delta\phi}$.
The term $L_4$ is necessary to determine the four-point expectation value, and
hence the corresponding trispectrum $T_{\delta\phi}$.
The scalar part of $L_4$ has already been derived by Sloth \cite{Sloth:2006az}.
In the following section we extend this result by incorporating the effect
of the vector part of $L_4$.

\subsection{The Lagrangian for field perturbations}

We consider the class of inflationary models where 
Einstein gravity is minimally coupled to
a set of scalar fields $\phi^\alpha$: 
\begin{equation}
\label{startingaction}
  S = \int \d t \, \d^3 x \; \sqrt{-g}\, \left( \frac{1}{2}R 
  - \frac{1}{2} \grad^a \phi^\alpha \grad_a \phi_\alpha - V \right)  ,
\end{equation}
where $g = \det g$, $R$ is the Ricci scalar of spacetime and 
$V$ is the interaction potential for the fields.
It is assumed that the potential can drive a phase of slow-roll inflation, but 
is otherwise left arbitrary. 

In order to work with the $\delta N$ formalism, we require the perturbations
$\delta\phi$ to be defined on `flat' spatial hypersurfaces on which the 
three-metric takes
the form $h_{ij} = a^2(t) [ \delta_{ij} + \exp(\gamma_{ij}) ]$, where
$\partial^i \gamma_{ij} = 0$. These slices are indeed flat in the conventional
sense if the tensor mode $\gamma_{ij}$ is absent, which for simplicity we
assume to be the case\footnote{On the basis of the calculations 
reported by Maldacena in
Ref.~\cite{Maldacena:2002vr}, any expectation
value involving a tensor mode will be smaller than an expectation value
involving only scalars. Since the \emph{scalar} non-gaussianity is already
hard to detect (and, despite a strong theoretical prejudice in its favour,
the tensor contribution has not yet even been confirmed
to exist), it seems reasonable to restrict attention to the scalar sector.
In that case, $\gamma_{ij}$ does not enter in the tree-level graphs and
can be dropped without loss of generality.}.
It proves convenient to employ the ADM form for the metric,
\begin{equation}
  \d s^2 = - N^2 \, \d t^2 
           + h_{ij} ( \d x^i + N^i \, \d t)
                    ( \d x^j + N^j \, \d t) ,
\end{equation}
where $N(t,\vect{x})$ and $N^i(t,\vect{x})$
denote the lapse function and shift
vector, respectively. 
The action \eref{startingaction} 
can be expressed in terms of these degrees of freedom 
such that\footnote{In what follows we employ a useful summation convention 
for the spatial indices, which are labelled $\{ i, j, \cdots \}$. 
When these appear as a raised/lowered
pair, it is to be understood that they are contracted with the 3-metric,
$h_{ij}$. On the other hand,  
if a pair appear with both indices lowered, they are contracted
with the flat Euclidean metric $\delta_{ij}$. In other words, 
\begin{equation}
  a^i b_i = \sum_{ij} h^{ij} a_i b_i \quad \mbox{and} \quad
  a_i b_i = \sum_{i} a_i b_i .
\end{equation} 
}
\begin{equation}
  \fl\label{action}
  S = - \frac{1}{2} \int \d t \, \d^3 x \; N \sqrt{h} \left(
  \grad^i \phi^\alpha \grad_i \phi_\alpha + 2 V \right) +
        \frac{1}{2} \int \d t \, \d^3 x \; \frac{\sqrt{h}}{N}
        \left( E^{ij} E_{ij} - E^2 + \pi^\alpha \pi_\alpha \right) ,
\end{equation}
where $E_{ij} = \frac{1}{2} \dot{h}_{ij} - \grad_{(i} N_{j)}$
is proportional to the extrinsic curvature of the spatial slices
and $\pi^\alpha =
\dot{\phi}^\alpha - N^j \grad_j \phi^\alpha$ is
proportional to the momentum conjugate to $\phi^\alpha$.

The degrees of freedom $N$ and $N^i$ may be integrated out by 
first extremizing the action with respect to these fields, 
solving the corresponding algebraic equations of 
motion, and then substituting the solutions back into the action. 
This results in an action that can be expressed entirely in terms 
of propagating fields. 
Although the solutions to the constraint equations are in general 
non-local, which is signalled (for example) by the presence of formal operators
such as the inverse Laplacian $\partial^{-2}$, this will not 
be problematic provided we remain interested only in momentum-space 
expectation values. On the other hand, 
the continuation of such expectation values back to real space may not
necessarily be a straightforward procedure. 

\subsubsection{The constraints}

The constraint equation for the lapse function is given by 
\begin{equation}
  \label{lapseconstraint}
  \grad^i \phi^\alpha \grad_i \phi_\alpha + 2 V +
  \frac{1}{N^2} \left( E^{ij} E_{ij} - E^2 + \pi^\alpha \pi_\alpha \right) = 0
\end{equation}
and the corresponding expression for the shift vector takes the form
\begin{equation}
  \label{shiftconstraint}
  \grad_i \left( \frac{1}{N} [E^i_j - \delta^i_j E] \right) =
  \frac{\pi^\alpha}{N} \grad_j \phi_\alpha .
\end{equation}

Eqs.~\eref{lapseconstraint} and~\eref{shiftconstraint}
can be solved perturbatively by writing
\begin{equation}
  N = 1 + \alpha,  \quad  \quad
  N_j = \grad_j \vartheta + \beta_j ,
\end{equation}
where $\beta_j$ is divergenceless, {\ie} $\grad_j \beta^j = 0$. It 
is then assumed that the variables $\alpha$, $\vartheta$ and $\beta_j$ 
can be expanded as a power series in the field perturbations:
\begin{equation}
  \alpha = \sum_{m = 1}^\infty \alpha_m , \quad  \quad
  \vartheta = \sum_{m = 1}^\infty \vartheta_m , \quad  \quad
  \beta_j = \sum_{m = 1}^\infty \beta_{mj} ,
\end{equation}
where a subscript $m$ denotes the number of powers of $\delta\phi$ that are
present in each term. As shown in Ref. \cite{Chen:2006nt}, 
although in principle we would need to include terms in the sum 
up to $m=4$ in order to obtain $L_4$, in practise the terms with $m=3$
and $m=4$ cancel out of the final answer.

Let us first consider the $N^i$ constraint. We find 
from the $\Or(\delta\phi)$ term that 		
\begin{equation}
  \alpha_1 = \frac{1}{2H} \dot{\phi}^\alpha \delta\phi_\alpha
  \quad \mbox{and} \quad
  \beta_{1j} = 0 ,
  \label{alpha1}
\end{equation}
whereas $\alpha_2$ and $\beta_{2j}$ are 
obtained from the $\Or(\delta\phi^2)$ term:
\begin{eqnarray}
  \fl\nonumber
  2H \partial_j \alpha_2 -
  \frac{1}{2a^2} \partial^2 \beta_{2j} =
  4H \alpha_1 \partial_j \alpha_1 +
  \delta \dot{\phi}^\alpha \partial_j \delta \phi_\alpha -
  \alpha_1 \dot{\phi}^\alpha \partial_j \delta \phi_\alpha \\
  \mbox{} + \frac{1}{a^2} \partial_i \alpha_1
  ( \delta_{ij} \partial^2 \vartheta_1 - \partial_i \partial_j
    \vartheta_1 )  .
  \label{beta}
\end{eqnarray}
In principle, the $\Or(\delta\phi^3)$ term could be used to 
find $\alpha_3$ and
$\beta_{3j}$, but these contributions are not needed.

With regard to the lapse function, the $\Or(1)$ term in the constraint 
equation \eref{lapseconstraint} is the Friedmann equation:
\begin{equation}
  6H^2 = 2V + \dot{\phi}^\alpha \dot{\phi}_\alpha .
\end{equation}
The $\Or(\delta\phi)$ term yields $\vartheta_1$,
\begin{equation}
  \frac{4H}{a^2} \partial^2 \vartheta_1 =
  -2 V_{,\alpha} \delta\phi^\alpha
  -2\dot{\phi}^\alpha \delta\dot{\phi}_\alpha +
  2\alpha_1(-6H^2 + \dot{\phi}^\alpha \dot{\phi}_\alpha) ,
  \label{theta1}
\end{equation}
and the term $\vartheta_2$ follows from the $\Or(\delta\phi^2)$ equation:
\begin{eqnarray}
  \fl\nonumber
  \frac{4H}{a^2} \partial^2 \vartheta_2 =
  - \frac{1}{a^2} \partial_i \delta\phi^\alpha \partial_i \delta\phi_\alpha -
  V_{,\alpha\beta} \delta\phi^\alpha \delta\phi^\beta \\\nonumber
  \mbox{} + \frac{1}{a^4} \partial^2 \vartheta_1 \partial^2 \vartheta_1 -
  \frac{1}{a^4} \partial_i \partial_j \vartheta_1 \partial_i \partial_j
  \vartheta_1 - \delta\dot{\phi}^\alpha \delta\dot{\phi}_\alpha +
  2 \frac{\dot{\phi}^\alpha}{a^2} \partial_i \vartheta_1 \partial_i
  \delta\phi_\alpha \\\label{theta2}
  \mbox{} + 2 \alpha_1 \left( \frac{4H}{a^2} \partial^2 \vartheta_1 +
  2 \dot{\phi}^\alpha \delta \dot{\phi}_\alpha \right) -
  (3 \alpha_1^2 - 2 \alpha_2)(-6H^2 + \dot{\phi}^\alpha \dot{\phi}_\alpha) .
\end{eqnarray}
Likewise, the $\Or(\delta\phi^3)$ term can be used to obtain $\vartheta_3$.
Note that since $\vartheta_2 \neq 0$ and it remains finite 
as $\dot{\phi}^{\alpha} \rightarrow 0$,
some metric back-reaction from the coupling to the scalar fields will still 
be present even in the extreme de Sitter limit.

The second-order perturbation theory of multiple scalar fields has
previously been studied by Malik in the large-scale limit
\cite{Malik:2005cy}.
Compared to the lower-order perturbation theory considered in
Refs.~\cite{Maldacena:2002vr,Seery:2005wm,Seery:2005gb},
the most distinctive new feature of the above analysis 
is the combination of $\alpha_2$ and $\beta_{2j}$
appearing in Eq.~\eref{beta}. These terms can be separated 
using the divergenceless condition on $\beta_j$ together with the 
reasoning that is employed in the proof of the
Helmholtz Theorem \cite{morse-feshbach,jeffreys-jeffreys}. For this
purpose, it is convenient to recall the elementary identity
\begin{equation}
  \label{vect-ident}
  \grad^2 \vect{W} = \gradient \divergence \vect{W} -
  \curl \curl \vect{W} ,
\end{equation}
for any vector field $\vect{W}$. 
A given vector $\vect{F}$ may be decomposed into its incompressible and
irrotational pieces by writing $\vect{F} = \gradient \varphi + \curl
\vect{A}$ and using Eq.~\eref{vect-ident} to identify
\begin{equation}
  \vect{A} = - \curl \frac{1}{\grad^2} \vect{F}
  \quad \mbox{and} \quad
  \varphi = \divergence \frac{1}{\grad^2} \vect{F} .
\end{equation}

After applying this construction to Eq.~\eref{beta}, 
it follows that $\alpha_2$ can be expressed in the form 
\begin{equation}
  \alpha_2 = \frac{\alpha_1^2}{2} + \frac{1}{2H} \partial^{-2}
  \left( \Sigma + \frac{1}{a^2} \partial^2 \alpha_1 \partial^2 \vartheta_1 -
  \frac{1}{a^2} \partial_i \partial_j \alpha_1 \partial_i \partial_j
  \vartheta_1 \right) ,
  \label{alpha2}
\end{equation}
where
\begin{equation}
  \Sigma \equiv \partial_j \delta \dot{\phi}^\alpha \partial_j \delta \phi
  _\alpha + \delta \dot{\phi}^\alpha \partial^2 \delta \phi_\alpha .
  \label{sigma}
\end{equation}
Likewise, it can be shown that to leading-order, 
$\beta_{2j}$ satisfies\footnote{The exact expression is
\begin{eqnarray}
  \fl\nonumber
  \frac{1}{2a^2} \beta_{2j} = \partial^{-4} \Big(
  \frac{1}{a^2} \partial^2 \alpha_1 \partial_j \partial^2 \vartheta_1
  -
  \frac{1}{a^2} \partial_m \partial_j \alpha_1 \partial_m \partial^2 \vartheta_1
  +
  \frac{1}{a^2} \partial_m \alpha_1 \partial_m \partial_j \partial^2 \vartheta_1
  -
  \frac{1}{a^2} \partial_j \alpha_1 \partial^4 \vartheta_1 \\ \nonumber
  \mbox{} -
  \frac{1}{a^2} \partial_m \partial_j \partial_i \alpha_1 \partial_i \partial_m \vartheta_1
  +
  \frac{1}{a^2} \partial^2 \partial_i \alpha_1 \partial_i \partial_j \vartheta_1
  -
  \frac{1}{a^2} \partial_j \partial_i \alpha_1 \partial_i \partial^2 \vartheta_1
  +
  \frac{1}{a^2} \partial_m \partial_i \alpha_1 \partial_m \partial_i \partial_j \vartheta_1 \\
  \mbox{} +
  \partial_m \partial_j \delta \dot{\phi}^\alpha \partial_m \delta \phi_\alpha
  -
  \partial^2 \delta \dot{\phi}^\alpha \partial_j \delta \phi_\alpha
  +
  \partial_j \delta \dot{\phi}^\alpha \partial^2 \delta \phi_\alpha
  -
  \partial_m \delta \dot{\phi}^\alpha \partial_m \partial_j \delta \phi_\alpha
  \Big) .
\end{eqnarray}
It is important that $\beta_{2j}$ \emph{cannot} be written as a total gradient.
Similar terms were identified in Refs. \cite{Rigopoulos:2005xx,Langlois:2006vv},
and in Ref. \cite{Langlois:2006vv} it was shown that such quantities decay like
$a^{-3}$ in an expanding universe. This is in agreement with
non-perturbative arguments which predict the decay of shear during inflation
\cite{wald-no-hair}, and is essential to maintain the relationship between
$\R$ and $\zeta$ on large scales.
We thank F. Vernizzi for emphasizing this point to us.}
\begin{equation}
  \fl\label{beta2}
  \frac{1}{2a^2} \beta_{2j} \simeq \partial^{-4} \left(
  \partial_m \partial_j \delta\dot{\phi}^\alpha \partial_m \delta\phi_\alpha +
  \partial_j \delta\dot{\phi}^\alpha \partial^2 \delta\phi_\alpha
  - \partial^2 \delta\dot{\phi}^\alpha \partial_j \delta\phi_\alpha -
  \partial_m \delta\dot{\phi}^\alpha \partial_m \partial_j \delta\phi_\alpha
  \right)
\end{equation}
where `$\simeq$' denotes equality only at leading-order in
the slow-roll expansion.

\subsubsection{The fourth-order action}

After use of the constraint equations \eref{lapseconstraint} 
and \eref{shiftconstraint} as outlined above, 
the fourth-order action is found to be given by   
\begin{eqnarray}
  \fl\nonumber
  S_4 = \int a^3 \Bigg( - \frac{1}{24} V_{,\alpha\beta\gamma\delta}
  \delta\phi^\alpha \delta\phi^\beta \delta\phi^\gamma \delta\phi^\delta
  + \frac{1}{2a^4} \partial_{(i} \beta_{2j)} \partial_i \beta_{2j} \\\nonumber
  \mbox{} + \frac{1}{2a^4} \partial_j \vartheta_1 \partial_j
  \delta\phi^\alpha \partial_m \vartheta_1 \partial_m \delta\phi_\alpha -
  \frac{1}{a^2} \delta\dot{\phi}^\alpha (\partial_j \vartheta_2 +
  \beta_{2j} ) \partial_j \delta\phi_\alpha \\\nonumber
  \mbox{} + (\alpha_1^2 \alpha_2 - \frac{1}{2}\alpha_2^2)
  (-6H^2 + \dot{\phi}^\alpha \dot{\phi}_\alpha) +
  \frac{\alpha_1}{2} \Big[ - \frac{1}{3} V_{,\alpha\beta\gamma}
  \delta\phi^\alpha \delta\phi^\beta \delta\phi^\gamma -
  2\alpha_1^2 V_{,\alpha} \delta\phi^\alpha \\\nonumber
  \quad\quad \mbox{} + \alpha_1 \Big( - \frac{1}{a^2} \partial_i
  \delta\phi^\alpha \partial_i \delta\phi_\alpha -
  V_{,\alpha\beta} \delta\phi^\alpha \delta\phi^\beta \Big) \\\nonumber
  \quad\quad \mbox{} - \frac{2}{a^4} \partial_i \partial_j \vartheta_2
  \partial_i \partial_j \vartheta_1 + \frac{2}{a^4} \partial^2
  \vartheta_2 \partial^2 \vartheta_1 - \frac{2}{a^4} \partial_i
  \beta_{2j} \partial_i \partial_j \vartheta_1 \\
  \quad\quad \mbox{} + \frac{2}{a^2} \dot{\phi}^\alpha
  (\partial_j \vartheta_2 + \beta_{2j}) \partial_j \delta\phi_\alpha +
  \frac{2}{a^2} \delta\dot{\phi}^\alpha \partial_j \vartheta_1
  \partial_j \delta\phi_\alpha \Big] \Bigg) .
  \label{foa}
\end{eqnarray}
When appropriately truncated in powers of slow-roll,
the scalar sector of this action is equivalent to that presented in
Appendix A of Ref. \cite{Sloth:2006az}. However no truncation in slow-roll
has been made in Eq.~\eref{foa}, which is exact.

The leading-order, slow-roll terms in Eq.~\eref{foa} arise from
$\alpha_2$, $\vartheta_2$, and $\beta_{2j}$
[via Eqs.~\eref{alpha2}, \eref{theta2} and \eref{beta2}]. 
These all contain terms of $\Or(\epsilon^0)$, where $\epsilon 
\equiv -\dot{H}/H^2$ parametrizes the degree of departure from a
pure de Sitter expansion. Extracting these contributions 
from Eq.~\eref{foa} then leads us to the fourth-order action 
for the perturbations that we have been seeking: 
\begin{equation}
  \fl\label{fourthorder}
  S_4 \simeq \int \d t \, \d^3 x \; \left(
  - \frac{1}{4a} \beta_{2j} \partial^2 \beta_{2j} +
  a \vartheta_2 \Sigma +
  \frac{3}{4} a^3 \partial^{-2} \Sigma \partial^{-2} \Sigma
  - a \delta \dot{\phi}^\alpha \beta_{2j}
  \partial_j \delta\phi_\alpha \right)
\end{equation}
The form of the lagrangian $L_4$ then follows immediately
from Eq.~\eref{fourthorder}. 

\subsection{Loop corrections}

The presence of interaction terms in $L_4$ 
which are `unprotected,' 
in the sense that they are not suppressed by a positive power of the slow-roll
parameters, might give rise to concerns that the two-point function
$\langle \delta\phi^\alpha \delta\phi^\beta \rangle$
also receives large, unprotected loop corrections. If this were the case, it
would seriously impair our ability to make 
predictions about the early universe from models of inflation.
Fortunately, however, the calculations reported in Ref.~\cite{Sloth:2006az} 
indicate that these unprotected loop corrections occur
at about 1\% of the tree-level, which is not large enough to harm
our predictivity. On the other hand, it should 
be remarked that it has yet to be verified 
that such corrections are \emph{never} large at any order in the
interaction of $\delta\phi$.

Given the interaction lagrangian \eref{fourthorder}, 
it only remains to apply Eqs.~\eref{ctp}--\eref{weight} 
in order to obtain the four-point expectation 
value of the field fluctuations 
and therefore the momentum-dependence 
of the form-factor $\Mfact_4$ and the non-linearity parameter $\tnl$. 
This will be focus of the following section. 

\section{The trispectrum form-factor $\Mfact_4$}
\label{sec:expect}

\subsection{The momentum-dependence of $\Mfact_4$}

It is convenient to calculate the four-point expectation value 
in three separate stages by considering the blocks of 
terms in Eq.~\eref{fourthorder} which consist of either two, one or zero 
copies of the pure vector $\beta_{2j}$.

\begin{itemize}
  \item \header{Vector $\times$ vector.} This is the term containing
  $\beta_{2j} \partial^2 \beta_{2j}$. It can be calculated following the
  general principles outlined 
  in \S\ref{sec:ff} and described in more detail in
  Refs.~\cite{Maldacena:2002vr,Weinberg:2005vy,Seery:2005wm,Seery:2005gb,
  Weinberg:2006ac,Sloth:2006az,Calzetta:1986ey,Jordan:1986ug,hajicek}.
  Before proceeding with the calculation, however, it will 
  be helpful to introduce some new notation.
  In particular, we define a vector $\Z{i}{j}$ which is a function 
  of two momenta $\vect{k}_i$ and $\vect{k}_j$, such that 
  \begin{equation}
    \Z{i}{j} = \sigma_{ij} \vect{k}_i - \sigma_{ji} \vect{k}_j
    \quad\mbox{(no summation)} ,
    \label{zk}
  \end{equation}
  where $\sigma_{ij}$ is the scalar combination
  \begin{equation}
    \sigma_{ij} = \vect{k}_i \cdot \vect{k}_j + k_j^2 .
    \label{sigmak}
  \end{equation}
  The vector $\times$ vector term then provides a contribution 
  to the four-point expectation value
  $\langle \delta\phi^\alpha(\eta',\vect{k}_1)
           \delta\phi^\beta(\eta',\vect{k}_2)
           \delta\phi^\gamma(\eta',\vect{k}_3)
           \delta\phi^\delta(\eta',\vect{k}_4) \rangle$ which is given by 
  \begin{eqnarray}
   \label{vectorvector}
    \fl\nonumber
    -\imag (2\pi)^3 \delta( \sum_i \vect{k}_i )
    \frac{H(\eta')^4}{16 \prod_i k_i^3}
    \left[ \prod_i (1 + \imag k_i \eta') \right] \e{-\imag \ktot \eta'}
    \delta^{\alpha\beta}
    \delta^{\gamma\delta}
    \frac{k_1^2 k_3^2}{k_{12}^2 k_{34}^4} \Z{1}{2} \cdot \Z{3}{4}
    \\
    \mbox{} \times \int_{-\infty}^{\eta_{\ast}} \d \eta \; H(\eta)^2
    (1 - \imag k_2 \eta) (1 - \imag k_4 \eta) \e{\imag \ktot \eta} +
    \mbox{permutations} + \mbox{c.c.},
  \end{eqnarray}
  where `c.c.' denotes the complex conjugate of the preceding term,
  the index $i$ ranges over $\{ 1, 2, 3, 4 \}$, 
  the conformal time equivalents of $( t' , t_\ast )$ 
  are given by $( \eta' , \eta_\ast )$ and
  the total scalar momentum is defined by $\ktot = k_1 + k_2 + k_3 + k_4$.
  The sum includes all permutations of the indices
  $\{ \alpha, \beta, \gamma, \delta \}$ which
  simultaneously permute the momentum labels $\{ 1, 2, 3, 4 \}$.
  
  The time of observation, $\eta'$, should be chosen to be a few e-foldings
  after the modes $\vect{k}_i$ have crossed the horizon, so that it can 
  be safely assumed that the fluctuations 
  $\delta\phi$ have become classical but have not undergone significant 
  evolution. At the level of accuracy to which
  we are working, this will be the case if the condition $|\ktot \eta'| \sim
  \Or(\epsilon) \ll 1$ is satisfied. When this condition holds,  
  we may further specify $H(\eta') = H(\eta) = H_{\ast}$ to a good 
  approximation, 
  since the integral over $\eta$ will receive only negligible contributions 
  from times long before the epoch of horizon
  crossing. Likewise, extending the upper limit
  $\eta_\ast$ in Eq.~\eref{vectorvector} 
  to the infinite future $(\eta_* \rightarrow 0^-)$ 
  introduces no significant error. This implies that 
  the integral \eref{vectorvector} can then be evaluated analytically and 
  we find that 
  the contribution of this piece to the form-factor $\Mfact_4$ is given by
  \begin{equation}
    \fl
    \Delta \Mfact_4 = - 2
    \frac{k_1^2 k_3^2}{k_{12}^2 k_{34}^4}
    \frac{W_{24}}{\ktot} \Z{1}{2} \cdot \Z{3}{4} ,
    \label{vv}
  \end{equation}
  where we have defined
  \begin{equation}
    \label{Wij}
    \fl
    W_{ij} = \imag \ktot \int_{-\infty}^0 \d \eta \;
      (1 - \imag k_i \eta)
      (1 - \imag k_j \eta) \e{\imag \ktot \eta}
    = 1 + \frac{k_i + k_j}{\ktot} +
                     \frac{2 k_i k_j}{\ktot^2} .
  \end{equation}
  Since Eq.~\eref{Wij} is purely real, the addition of its 
  complex conjugate has been absorbed into an overall factor of 2.
  
  \item \header{Scalar $\times$ vector.}
  The same principles apply to the scalar $\times$ vector sector, 
  which comprises the term
  $a \delta\dot{\phi}^\alpha \beta_{2j} \partial_j \delta\phi_\alpha$.
  Its contribution to the four-point expectation value is given by 
  \begin{eqnarray}
    \fl\nonumber
    -\imag (2\pi)^3 \delta(\sum_i \vect{k}_i) \frac{H_{\ast}^6}{16
    \prod_i k_i^3} 2 \delta^{\alpha\beta} \delta^{\gamma\delta}
    \frac{k_1^2 k_3^2}{k_{34}^4}
    \vect{k}_2 \cdot \Z{3}{4} \\
    \mbox{} \times \int_{-\infty}^0 \d \eta \; (1 - \imag k_2 \eta)
    (1 - \imag k_4 \eta) \e{\imag \ktot \eta}
    + \mbox{permutations} + \mbox{c.c.}
  \end{eqnarray}
  Since the integral is equivalent to $W_{24}$, we may immediately deduce the
  corresponding contribution of this term to the form-factor is 
  \begin{eqnarray}
    \fl \Delta \Mfact_4 = - 4 \frac{k_1^2 k_3^2}{k_{34}^4}
    \frac{W_{24}}{\ktot} \vect{k}_2 \cdot \Z{3}{4} .
    \label{sv}
  \end{eqnarray}
  
  \item \header{Scalar $\times$ scalar.}
  The final piece is comprised of $a \vartheta_2 \Sigma + (3/4) a^3
  \partial^{-2} \Sigma \partial^{-2} \Sigma$ and is independent of $\beta_i^2$. 
  A loop integral involving this term was calculated in 
  Ref.~\cite{Sloth:2006az}.
  
  After truncating $\vartheta_2$ to leading order, using Eq.~\eref{theta2},
  the scalar piece can be rewritten
  \begin{equation}
    \fl
    S_{\mathrm{scalar}} = \int \d t \, \d^3 x \;
    \left[ \frac{a^3}{4H} \partial^{-2} \Sigma
    \left( - \frac{1}{a^2} \partial_i \delta \phi^\alpha
    \partial_i \delta \phi_\alpha - \delta\dot{\phi}^\alpha
    \delta\dot{\phi}_\alpha \right) -
    \frac{3}{4} a^3 \partial^{-2} \Sigma \partial^{-2} \Sigma \right] .
  \end{equation}
  The first of these terms makes a contribution of the form
  \begin{eqnarray}
  \label{firstscalarscalar}
    \fl\nonumber
    -\imag (2\pi)^3 \delta(\sum_i \vect{k}_i) \frac{H_{\ast}^6}{16
    \prod_i k_i^3} \frac{1}{4} \delta^{\alpha\beta} \delta^{\gamma\delta}
    \frac{k_3^2}{k_{34}^2} \sigma_{34} \\ \fl \nonumber
    \mbox{} \times \int_{-\infty}^{0} \d \eta \; \left[
    \vect{k}_1 \cdot \vect{k}_2
    (1 - \imag k_1 \eta)(1 - \imag k_2 \eta)(1 - \imag k_4 \eta) -
    k_1^2 k_2^2 \eta^2 (1 - \imag k_4 \eta) \right] \e{\imag \ktot \eta} \\
    \mbox{} + \mbox{permutations} + \mbox{c.c.}
  \end{eqnarray}
  It proves useful when evaluating this integral to define a
  \emph{three}-momentum generalization of $W_{ij}$,
  \begin{eqnarray}
    \fl\nonumber
    W_{\ell m n} = \imag \ktot
    \int_{-\infty}^{0} \d \eta \;
    (1- \imag k_\ell \eta) (1 - \imag k_m \eta) (1 - \imag k_n \eta)
    \e{\imag \ktot \eta} \\
    \lo{=}
      1 + \frac{k_\ell + k_m + k_n}{\ktot} +
      \frac{2(k_\ell k_m + k_\ell k_n + k_m k_n)}{\ktot^2} +
      \frac{6 k_\ell k_m k_n}{\ktot^3} .
  \end{eqnarray}
  Hence, the contribution of Eq.~\eref{firstscalarscalar} to the form-factor 
  is
  \begin{equation}
    \fl \Delta \Mfact_4 = - \frac{1}{2} \frac{k_3^2}{k_{34}^2} \sigma_{34}
    \sigma_{34} \left[ \frac{\vect{k}_1 \cdot \vect{k}_2}{\ktot}
    W_{124} + \frac{k_1^2 k_2^2}{\ktot^3} \left(
    2 + 6 \frac{k_4}{\ktot} \right) \right] .
    \label{ss1}
  \end{equation}

  Finally, the second term, $-(3/4) a^3
  \partial^{-2} \Sigma \partial^{-2} \Sigma$, yields
  \begin{eqnarray}
    \fl\nonumber
    -\imag (2\pi)^3 \delta(\sum_i \vect{k}_i) \frac{H_{\ast}^6}{16\prod_i k_i^3}
    \frac{3}{4} \delta^{\alpha\beta}\delta^{\gamma\delta}
    \frac{k_1^2 k_3^2}{k_{12}^2 k_{34}^2}
    \sigma_{12} \sigma_{34} \\
    \times \int_{-\infty}^0 \d \eta \;
    (1 - \imag k_2 \eta)(1 - \imag k_4 \eta) \e{\imag \ktot \eta}
    + \mbox{permutations} + \mbox{c.c.}
  \end{eqnarray}
  and provides the contribution
  \begin{equation}
    \Delta \Mfact_4 =
    -\frac{3}{2} \frac{k_1^2 k_3^2}{k_{12}^2 k_{34}^2}
    \sigma_{12} \sigma_{34} \frac{W_{24}}{\ktot}
    \label{ss2}
  \end{equation}
  to the form-factor. 
\end{itemize}

It follows, therefore, that the complete momentum-dependence of the 
trispectrum can be determined by assembling Eqs.~\eref{vv},
~\eref{sv}, ~\eref{ss1} and ~\eref{ss2}. Specifically, 
we find that the form-factor is given by  
\begin{eqnarray}
  \fl\nonumber
  \Mfact_4 = - 2 \frac{k_1^2 k_3^2}{k_{12}^2 k_{34}^2}
  \frac{W_{24}}{\ktot}
  \left[
  \frac{\Z{1}{2} \cdot \Z{3}{4}}{k_{34}^2}
  + 2 \vect{k}_2 \cdot \Z{3}{4}
  + \frac{3}{4} \sigma_{12} \sigma_{34} \right] \\
  \mbox{} - \frac{1}{2} \frac{k_3^2}{k_{34}^2} \sigma_{34}
  \left[\frac{\vect{k}_1 \cdot \vect{k}_2}{\ktot} W_{124} +
        \frac{k_1^2 k_2^2}{\ktot^3} \left( 2 + 6 \frac{k_4}{\ktot} \right)
  \right] ,
  \label{mfour}
\end{eqnarray}
and Eq.~\eref{mfour} represents the main result of the paper.
The form-factor has dimensions of three powers of the momentum, 
which is the same behaviour as the corresponding form-factor $\Mfact_3$
which determines the bispectrum. However, the momentum-dependence 
of $\Mfact_4$ is considerably more complicated than 
that of $\Mfact_3$ and, in particular, it depends on the products
$\vect{k}_i \cdot \vect{k}_j$ and not merely the magnitudes $k_i$.

\subsection{The limit of equal momentum}

If two of the momenta, say $\vect{k}_1$ and $\vect{k}_2$, 
have equal magnitude $k_1 = k_2 \equiv p$, and are also equal and opposite,
$\vect{k}_1 + \vect{k}_2 = 0$, it follows from  
momentum conservation that the other two momenta must also 
be equal and opposite, albeit with a possibly different common magnitude, {\ie}, 
$k_3 = k_4 \equiv q$. On such configurations,
$\Mfact_4$ appears to be pathological, since the factors
$(\vect{k}_1 + \vect{k}_2)^{-1}$ and $(\vect{k}_3 + \vect{k}_4)^{-2}$
na\"{\i}vely diverge. If this were the case it would be disastrous,
since the non-gaussianity associated with the four-point 
function is presumably bounded from above by the WMAP data. 

Despite appearances,
the form-factor $\Mfact_4$ does not diverge on these degenerate 
quadrilaterals. This can be verified by introducing 
a simple specification for the $\vect{k}_i$. 
We define a matrix
of angles $\theta_{ij} \in [0,\pi]$ which satisfy
\begin{equation}
  \cos \theta_{ij} = \frac{\vect{k}_i \cdot \vect{k}_j}{k_i k_j} 
  \quad \mbox{for $i \neq j$}.
\end{equation}
This matrix obeys the obvious symmetry constraint $\theta_{ij} =
\theta_{ji}$, and therefore has 6 components, which cannot
all be chosen independently. To fully specify
the $\{ \vect{k}_i \}$, it is necessary to include two
additional angles which determine the
absolute orientation of
one of the $\vect{k}_i$, which we take to be $\vect{k}_4$.
However, these angles do not contribute
to $\Mfact_4$ and we need not be explicit about their
precise assignment. There is another angular degree of freedom
corresponding to rigid rotations which leave $\vect{k}_4$ fixed,
which does not change the intrinsic geometry of the $\{ \vect{k}_i \}$.
In total, therefore, the magnitudes of the four momenta, 
the two angles which determine the orientation of $\vect{k}_4$,
the angular degree of freedom corresponding to rotations which leave
$\vect{k}_4$ fixed,
and the five independent
$\theta_{ij}$ comprise the twelve degrees of freedom which are  
required to completely specify four vectors in $\reals^3$.

Momentum conservation supplies three constraints which eliminate three of
these degrees of freedom. These can be chosen to be three of the angles
$\theta_{ij}$.
It is convenient to work with the angles
$\{ \theta_r  = \theta_{r4} \}$, in terms of which the others can be written as
\begin{equation}
  \cos \theta_{13} = \cos \theta_{2} \label{m1} ,
\end{equation}
\begin{equation}
  \cos \theta_{23} = \cos \theta_{1} \label{m2} ,
\end{equation}
\begin{equation}
  \cos \theta_{12} = \frac{q^2}{p^2}(1 + \cos \theta_3) - 1 \label{m3} .
\end{equation}

We wish to re-express the form-factor \eref{mfour} in terms of 
$p$, $q$ and $\theta_r$.  
This requires explicit expressions for
$W_{24}$ and $W_{124}$, which can be shown to be  
\begin{eqnarray}
  \label{usefulone}
  W_{24} = \frac{6(p+q)^2 + 2pq}{4(p+q)^3} \label{W} \\
  W_{124} = \frac{8(p+q)^3 + 4(p+q)^2(2p+q) + 4p(p+q)(2q+p) + 6p^2q}
                      {8(p+q)^3} \label{Gamma} .
\end{eqnarray}
We also have the additional relation 
$\sigma_{12} = \sigma_{34} = q^2(1 + \cos \theta_3)$ and the quantities  
$\Z{1}{2} \cdot \Z{3}{4}$ and $\vect{k}_2 \cdot \Z{3}{4}$ can be
reduced to  
\begin{eqnarray}
  \Z{1}{2} \cdot \Z{3}{4} = 2 p q^5(1 + \cos \theta_3)^2(\cos \theta_2 -
                                                         \cos \theta_1)
                                                         \label{z1} \\
  \vect{k}_2 \cdot \Z{3}{4} = - p q^3(1 + \cos \theta_3)(\cos \theta_2 -
                                                         \cos \theta_1) ,
                                                         \label{z2}
\end{eqnarray}
respectively. 

It is now relatively straightforward to rewrite expression 
\eref{mfour} given Eqs.~\eref{usefulone}--\eref{z2} and we 
find after some algebra that  
\begin{eqnarray}
  \fl\nonumber
  \Mfact_4 = \frac{W_{24}}{4(1+\cos\theta_3)} \frac{p^2 q}{p+q} \left[
    p(\cos\theta_2 - \cos\theta_1) - \frac{3q}{4}(1 + \cos\theta_3) \right] \\
  \mbox{} - \frac{q^2 p^2}{8(p+q)} \left[
    W_{124} \left\{ \frac{q^2}{p^2}(1 + \cos \theta_3) - 1 \right\}
    + \frac{p^2(2p+5q)}{4(p+q)^3} \right] .
  \label{mfourpq}
\end{eqnarray}
The na\"{\i}ve divergence associated with the limit
$\vect{k}_1 \rightarrow - \vect{k}_2$ reappears 
as $\theta_3 \rightarrow \pi$. However, Eq.~\eref{mfourpq}
should not be considered
in isolation, but only after symmetrization over the labels
$\{ 1, 2, 3, 4 \}$.
Indeed, after symmetrization over the exchange
$1 \leftrightharpoons 2$ we find 
\begin{equation}
  \fl
  \Mfact_4(1 \leftrightharpoons 2)
           = -\frac{3 p^2 q^2}{8(p+q)} W_{24} - \frac{p^2 q^2}{4(p+q)}
             \left[ W_{124} \left\{ \frac{q^2}{p^2}(1+\cos\theta_3)-1
             \right\} + \frac{p^2(2p+5q)}{4(p+q)^3} \right] .
  \label{finite}
\end{equation}
and it follows that Eq.~\eref{finite} 
possesses a finite limit as $\theta_3 \rightarrow \pi$ which
is not spoilt by further symmetrization.
Hence, as anticipated, Eq.~\eref{finite} 
demonstrates that $\Mfact_4$ does \emph{not} diverge on
degenerate quadrilaterals.

\section{How large can the trispectrum be?}
\label{sec:bound}

We now proceed to estimate the magnitude of the non-linearity 
parameter $|\Delta \tnl|$ that arises from
Eq.~\eref{mfour}. This is important, since the non-gaussianity associated
with the trispectrum
may be observable in the near future if
$|\Delta\tnl| \gtrsim 560$ \cite{Kogo:2006kh}. 

The form-factor $\Mfact_4$ is formally valid only when
$k_i \approx k_j$ for all $i$ and $j$.
We will show in \S\ref{sec:consistency} that, at least in the single-field
case, $\Mfact_4$ must become of order $\Or(\epsilon)$ in the limit where one
$\vect{k}_i$ becomes much smaller than the others. In this limit
one cannot trust \eref{mfour} to give the dominant contribution to
$\Mfact_4$, but it is consistent
with the explicit momentum dependence in \eref{mfour}, which approaches zero
in the formal limit where any $k_i \rightarrow 0$. This provides evidence
that the largest contribution to $\tnl$ from the connected four-point
function of the $\{ \delta\phi^\alpha \}$ arises on equilateral configurations,
although a more complete treatment of this question would be desirable.
In this section, we assume that $|\Delta\tnl|$ \emph{is} maximized
on equilateral configurations of the $\{ \vect{k}_i \}$.

The relevant expressions for the equilateral configuration 
follow after we have specified $p = q = k$ in Eqs.~\eref{m1}--\eref{z2}.
The equilateral condition is \emph{not} independent of momentum
conservation. In practise this means that the angles $\theta_r$ cannot
be chosen freely, but must instead obey the constraint
\begin{equation}
  \sum_{r} \cos \theta_r = -1 .
  \label{equilateral}
\end{equation}
We may choose only two of the $\theta_r$ arbitrarily.
These two angles suffice to parametrize the family of
distinct equilateral configurations. If $\vect{k}_4$ is
chosen to point along $\hat{\vect{z}}$, and we take
\eref{equilateral} to determine $\theta_3$ in terms of
$\{ \theta_1, \theta_2 \}$, then this family can be written
\begin{equation}
  \left. \begin{array}{l}
    \vect{k}_1 = (\sin \theta_1 \cos \phi_1,
                  \sin \theta_1 \sin \phi_1,
                  \cos \theta_1) \\
    \vect{k}_2 = (\sin \theta_2 \cos (\phi_1 + \alpha),
                  \sin \theta_2 \sin (\phi_1 + \alpha),
                  \cos \theta_2) \\
    \vect{k}_3 = (\sin \theta_3 \cos (\phi_1 + \beta),
                  \sin \theta_3 \sin (\phi_1 + \beta),
                  \cos \theta_3) \\
    \vect{k}_4 = (0,0,1)
  \end{array} \right\} ,
  \label{kconfig}
\end{equation}
where $\cos \theta_3 = -1 - \cos \theta_1 - \cos \theta_2$,
and the angles $\alpha$ and $\beta$ obey
\begin{equation}
  \tan \alpha = \pm \frac{\sqrt{1 - \gamma_3^2}}{\gamma_3},
  \quad \mbox{and} \quad
  \tan \beta  = \mp \frac{\sqrt{1 - \gamma_2^2}}{\gamma_2} ,
\end{equation}
given that $\gamma_m$ is a combination of the $\theta_m$,
\begin{equation}
  \gamma_m = \frac{\cos \theta_m - \prod_{r \neq m} \cos \theta_r}
                  {\prod_{s \neq m} \sin \theta_s} .
\end{equation}
The extra degree of freedom, $\phi_1$, corresponds to
rigid rotations of the $\{ \vect{k}_1, \vect{k}_2, \vect{k}_3 \}$
which fix $\vect{k}_4 \propto \hat{\vect{z}}$.
One can verify explicitly that Eqs.~\eref{m1}--\eref{m3} and
Eq.~\eref{equilateral} are obeyed by \eref{kconfig},
irrespective of the value of $\phi_1$.
In the equilateral limit, $W_{24}$ and $W_{124}$ reduce to
\begin{eqnarray}
  W_{ij} \rightarrow \frac{13}{8} \\
  W_{\ell m n} \rightarrow \frac{71}{32} ,
\end{eqnarray}
respectively, and this implies that Eq.~\eref{mfourpq} simplifies to 
\begin{equation}
  \fl
  \Mfact_4 = 
  \frac{k^3}{4} \left[
  \frac{13}{16} \frac{1}{1+\cos\theta_3} \left\{ \cos \theta_2 - \cos \theta_1
  - \frac{3}{4} ( 1 + \cos \theta_3 ) \right\} - \frac{71}{128} \cos \theta_3
  - \frac{7}{128} \right] .
\end{equation}
We may then substitute this result into Eq.~\eref{tnlr}.
Summing over the possible permutations of labels
and using Eq.~\eref{equilateral} to eliminate the sum $\sum_r \cos \theta_r$,
we deduce that
\begin{equation}
  \Delta \tnl = - \frac{23\sqrt{2}r}{32}
  \left[ \sum_r (1+\cos\theta_r)^{-3/2} \right]^{-1}.
  \label{finaltnl}
\end{equation}
The magnitude of $\tnl$ which is generated depends on the angles
$\{ \theta_r \}$. However, this dependence is in some sense artificial
since the totally permuted form factor, $\sum_{\mathrm{perms}} \Mfact_4$,
contains no angular dependence in the equilateral limit. The dependence on
the $\{ \theta_r \}$ which appears in Eq.~\eref{finaltnl} arises solely
from the definition of $\tnl$ which we adopted in Eq.~\eref{tnl}.
$|\Delta\tnl|$ is maximized on the configuration $\cos \theta_r = -1/3$,
which gives
\begin{equation}
  |\Delta \tnl| < \frac{23}{576\sqrt{3}} r \simeq 0.023 r
  \sim \frac{r}{50} .
  \label{toosmall}
\end{equation}
Since the current experimental upper bound on the tensor-to-scalar ratio is
$r < 0.55$ (at 95\% confidence) \cite{Spergel:2006hy}, 
Eq.~\eref{toosmall} immediately implies 
that the primordial non-gaussianity in the trispectrum 
that is generated at horizon crossing will be unobservably small. 
Consequently, as in the case of the bispectrum, 
if a non-trivial primordial trispectrum is observed in a future CMB experiment,  
its origin is unlikely to arise from quantum-mechanical interference around
the time of horizon crossing.

We observe that \eref{toosmall} is not zero. Taken together with the explicit
family of equilateral configurations \eref{kconfig}, which are all allowed by
momentum conservation, this removes any worry that the complicated expression
obtained for $\Mfact_4$ in Eq.~\eref{mfour} might be constrained to zero
by purely kinematical considerations.

Thus far, our discussion has considered the multi-field inflationary
scenario. A number of simplifications arise when the 
analysis is restricted to single-field inflation and we
proceed to discuss these in the following section. 

\section{Single-field Inflation
and the Maldacena consistency relation}
\label{sec:consistency}

In single field inflation, $\zeta$ is conserved to all orders
on superhorizon scales
\cite{Wands:2000dp,Lyth:2004gb}. This has the immediate consequence
that no evolution in $\tnl$ is possible after horizon exit,
and Eq.~\eref{toosmall} represents the largest possible non-gaussian
signal which can be visible in the trispectrum. We conclude that it is
too small ever to be detected.

A second special feature of single-field inflation is the existence of
a consistency relation between the correlators
$\langle \zeta(\vect{k}_1) \cdots \zeta(\vect{k}_n) \rangle$. In the
`squeezed' limit, where $k_1 \rightarrow 0$, the $\zeta(\vect{k}_1)$
mode crosses the horizon at a much earlier epoch than 
the remaining modes $\{ \vect{k}_2, \ldots, \vect{k}_n \}$.
By the time these modes eventually exit the horizon, the
gravitational background has been deformed due to the presence of the
$\zeta(\vect{k}_1)$ mode. As first pointed out by Maldacena 
\cite{Maldacena:2002vr}, the $\vect{k}_1$ mode and the
$\{ \vect{k}_2, \cdots, \vect{k}_n \}$ modes no longer interfere
when they cross the horizon, which implies that
the only correlation between them is the one imposed by this gravitational
relationship. In the case of the three-point function, this yields
the relation
\begin{equation}
  \langle \zeta(\vect{k}_1) \zeta(\vect{k}_2) \zeta(\vect{k}_3) \rangle
  \rightarrow - (n_s-1) P_\zeta(k_1)
  \langle \zeta(\vect{k}_2) \zeta(\vect{k}_3) \rangle
  \quad
  \mbox{(as $k_1 \rightarrow 0$)} ,
  \label{malda-consist}
\end{equation}
where $n_s$ is the spectral index of the scalar perturbation 
power spectrum. 
It was later emphasized by Creminelli \& Zaldarriaga \cite{Creminelli:2004yq}
that the limit~\eref{malda-consist} is purely kinematical and applies 
for any metric theory of gravity. This implies that it can be calculated 
on the basis of gravitational physics alone \cite{Allen:2005ye}.

Maldacena's arguments apply for any correlation function of $\zeta$, 
including the four-point function. In this case, the consistency
relationship is given by
\begin{equation} 
  \fl
  \langle \zeta(\vect{k}_1) \zeta(\vect{k}_2) \zeta(\vect{k}_3)
          \zeta(\vect{k}_4) \rangle
  \rightarrow - P(k_1) \frac{\d}{H \, \d t}
  \langle \zeta(\vect{k}_2) \zeta(\vect{k}_3) \zeta(\vect{k}_4) \rangle
  \quad
  \mbox{(as $k_1 \rightarrow 0$)} .
  \label{fourpt-consist}
\end{equation}

To gain further insight into the nature 
of Eq.~\eref{fourpt-consist}, it is instructive to count powers 
of slow-roll. 
In general, the derivatives of the number of e-foldings, $N$, 
are controlled by the slow-roll parameter $\epsilon = - \dot{H}/H^2$. If 
$\epsilon$ is almost constant over the range of e-folds
under consideration, it follows that $N \simeq - \epsilon^{-1}
\ln H$. This implies that 
each derivative of $N$ with respect to the scalar field 
generates an extra power of $\epsilon^{1/2}$, {\ie}, 
$N_{,\phi} \sim \Or(\epsilon^{-1/2})$,
$N_{,\phi\phi} \sim \Or(1)$ and
$N_{,\phi^n} \sim \Or(\epsilon^{(n-2)/2})$, etc.
By counting powers of slow-roll in Eq.~\eref{fourpt-consist}, 
it may be verified that this behaviour produces a four-point function of order
$\Or(\epsilon^{-1})$. This can be compared to 
the horizon-crossing contribution, Eq.~\eref{fourpt-a}, which is
$\Or(\epsilon^{-2})$.

Eq.~\eref{fourpt-consist} implies that \eref{mfour} becomes at most
of $\Or(\epsilon)$ when any of the $k_i \rightarrow 0$, at least in the
single field case.
Therefore, the leading slow-roll term in
the four-point function of $\zeta$ will be given by possible
subdominant corrections to \eref{mfour} which do not vanish
in the squeezed limit, together with the terms
which contribute to the four-point function 
at \emph{next}-order in the slow-roll
parameter $\epsilon$, {\ie}, the terms of order $\Or(\epsilon^{-1})$.

The first such term is
\begin{equation}
  \fl
  (2\pi)^3 \delta( \sum_i \vect{k}_i ) \frac{1}{2} N_{,\phi\phi} N_{,\phi}
  N_{,\phi} N_{,\phi}
  \left[ P(k_2) B(k_{12},k_3,k_4) + \mbox{23 permutations} \right] ,
  \label{malda1}
\end{equation}
where the sum contains twenty-four terms that are
obtained by rearrangements of the momenta, and $B$ is the
$\delta\phi$-bispectrum. The momentum conservation condition
implies that one-half of the permuations are equal to the other half,
giving twelve distinct terms.

The second relevant term is
\begin{equation}
  \fl
  (2\pi)^3 \delta( \sum_i \vect{k}_i ) \frac{1}{2} N_{,\phi\phi} N_{,\phi\phi}
  N_{,\phi} N_{,\phi}
  \left[ P(k_{13}) P(k_3) P(k_4) + \mbox{23 permutations} \right] .
  \label{malda2}
\end{equation}
At the tree-level, only this term was included in the estimate
for $\tnl$ given by
Alabidi \& Lyth \cite{Alabidi:2005qi} and Lyth \cite{Lyth:2006gd},
although these authors also included a one-loop diagram whose contribution
we assume to be negligible in comparison with the tree-level.%
\footnote{Owing to a typographical error, the $\tnl$ which follows from
\eref{malda2} appeared incorrectly in the journal version of
Ref. \cite{Lyth:2006gd}. The correct expression was given in both
\texttt{v1} and \texttt{v2} of Ref. \cite{Alabidi:2005qi}, available from
the arXiv. [D. Lyth, private communication (2006).]}

The final contribution at $\Or(\epsilon^{-1})$ is
\begin{equation}
  \fl
  (2\pi)^3 \delta( \sum_i \vect{k}_i ) \frac{1}{6} N_{,\phi\phi\phi}
  N_{,\phi} N_{,\phi} N_{,\phi} \left[ P(k_2) P(k_3) P(k_4) +
  \mbox{23 permutations} \right] .
  \label{malda3}
\end{equation}
An expression equivalent to this was given in
Ref. \cite{Sasaki:2006kq} in the context of the curvaton scenario.

In principle, therefore, the next-order contribution to the 
four-point function can be obtained by 
summing Eqs.~\eref{malda1}--\eref{malda3}. However, when performing such 
a sum, self-consistency would require that the next-order
versions of the two- and three-point
functions of the field fluctuations $\delta \phi$ and 
the derivatives of $N$ be employed. To date, the required expressions for 
these quantities have yet to be calculated 
since presumably they are unobservably small. Consequently, 
we should not expect the $\epsilon$ and $\eta$
dependence of Eqs.~\eref{malda1}--\eref{malda3} to match the right-hand
side of Eq.~\eref{fourpt-consist}, where
$\eta = V_{,\phi\phi}/V$ is $\Or(\epsilon)$
in the slow-roll expansion. On the other hand, the next-order
corrections to the scale factor, $a(t)$, 
the wavefunctions of the $\delta\phi$ and the
third-order action are not expected to contain any
\emph{intrinsically}
second- or higher-order slow-roll parameters, which involve
the third- or higher-derivatives
of the potential. Therefore, summing Eqs.~\eref{malda1}--\eref{malda3}
with the leading-order expressions for $N$,
the scale factor $a$, and the two- and three-point
functions of the field flucations $\delta \phi$ should match the
coefficient of $V_{,\phi\phi\phi}$ in
Eq.~\eref{fourpt-consist}. We have verified that this is the case.

\section{Conclusions}
\label{sec:conclude}
In this paper, we have derived the trispectrum of the 
primordial curvature perturbation
imprinted at horizon crossing during inflation.
In particular, we have estimated the expected magnitude of the
non-linearity parameter, $\tnl$, which is generated by this process.
We find that it is bounded by about $r/50$, and
therefore is too small to be observable 
by the Planck satellite or other future CMB experiments.
In single-field models,
$\zeta$ is conserved to all orders on superhorizon scales.
Therefore, no subsequent generation of non-gaussianity
is possible, and the primordial signal will be too small
ever to be detected.

One unexpected feature of the four-point action is that it contains no
powers of slow-roll parameters. In an earlier publication
\cite{Seery:2005gb}, two of us (DS and JEL)
suggested that the $n$th order action would contain essentially
$\lfloor n/2 \rfloor$ powers of $\dot{\phi}/H$, which would imply that
the coupling to the spacetime metric (and all backreaction from it)
switches off
order-by-order in the limit $\dot{\phi}/H \rightarrow 0$, and therefore
that the field fluctuations are purely Gaussian in the de Sitter limit.
Our result for $\Mfact_4$ shows that this does \emph{not} happen:
the scalar fields and the metric remain coupled even in exact de Sitter
space, and the fluctuations are therefore non-gaussian.
Intuitively, this occurs because the size of the
non-gaussianities in the $\delta\phi$ is
controlled by the strength of the gravitational interaction,
rather than $\dot{\phi}/H$, even though powers of slow-roll appear once one
has changed variable to the comoving curvature perturbation, $\zeta$.
We note, however, that $\zeta$ is not an appropriate choice of
variable to discuss the limit $\dot{\phi}/H \rightarrow 0$, since it
becomes singular there.

In the case of single-field inflation, we
have applied Maldacena's consistency argument to the trispectrum.
For the first time, this has employed the \emph{third}-order term 
in the expansion of $\delta N$
in terms of initial field values. 
Only the linear and quadratic terms in
the expansion have previously been required.
In the case of the trispectrum, the consistency relationship arises as a 
next-order effect in the slow-roll parameter $\epsilon$, 
in contrast to that of the bispectrum. 

We have also derived expressions in the spatially flat gauge 
for the second-order metric perturbation sourced by multiple scalar
fields.
These expressions
show explicitly that a
metric perturbation continues to exist at this order even when the
background is pure de Sitter space, corresponding to 
$\epsilon \rightarrow 0$. At first-order, the metric perturbation
disappears in this limit.

\ack
DS is supported by PPARC grant PPA/G/S/2003/00076.
We thank D. Lyth and the
Department of Physics at Lancaster University for their hospitality during
the workshop \emph{Non-gaussianity from Inflation}, June 2006.
We also thank K. Malik, D. Lyth and F. Vernizzi
for useful discussions and comments on
earlier versions of this paper.

\section*{References}
\providecommand{\href}[2]{#2}\begingroup\raggedright\endgroup

\end{document}